# Impact of sintering temperature on room temperature magneto-resistive and magneto-caloric properties of $Pr_{2/3}Sr_{1/3}MnO_3$

Ramesh Chandra Bhatt[1,2], Shiva Kumar Singh[1], P. C. Srivastava[2], S.K.Agarwal[1] and V. P. S. Awana[1,*]

[1]Quantum Phenomena and Applications Division, National Physical Laboratory (CSIR), New Delhi-110012, India

[2]Department of Physics, Banaras Hindu University, Varanasi-221005, India

## Abstract

Magneto-resistive and magneto-caloric properties of polycrystalline $Pr_{2/3}Sr_{1/3}MnO_3$ have been studied as a function of sintering temperature ($T_s$) between 1260-1450°C. Reitveld refinement of their X-ray diffraction (*XRD*) patterns confirms their single phase crystalline structure with orthorhombic *Pbnm* space group. The point of maximum value of temperature coefficient of resistance ($TCR_{max}$) and Curie temperature ($T_c$) decreased slightly with $T_s$. Magneto-resistance (*MR*) and magnetic entropy ($\Delta S_M$) increased markedly with sintering temperature. This could be attributed to the observed sharpness of both the magnetic and resistive transitions due to better grain connectivity. Optimum results are obtained for the sample with $T_s$ = 1400°C. *MR* at $T_c$ of the same is found to be as large as 32% at 1T and 58% at 5T magnetic fields. The maximum entropy change ($\Delta S_M max$) near its $T_c$ is 2.3Jkg$^{-1}$K$^{-1}$ and 7.8 Jkg$^{-1}$K$^{-1}$ upon 1T and 5T fields change respectively. These characteristics [*MR* (32% 1T, 58% 5T) and reasonable change in magnetic entropy (7.8Jkg$^{-1}$K$^{-1}$, 5T)] generate possibility that the optimized compound can be used as a potential magnetic refrigerant close to room temperature.

*Corresponding Author awana@mail.nplindia.ernet.in

Tel.: +91 11 45609357; Fax; +91 11 45609310

*Web page*- www.freewebs.com/vpsawana/

## 1. Introduction

Colossal magneto-resistive (*CMR*) manganese oxides having general formula $R_{1-x}A_xMnO_3$ (R is trivalent rare-earth cation and A is divalent alkaline-earth cation) have been considered to be potential candidates for wide range of applications like magnetic sensors, bolometric devices and magnetic refrigeration [1-8]. Properties which lead them as promising candidates for application are their higher temperature coefficient of resistance (*TCR*), *CMR* and magnetic entropy change near room temperature. Mostly these manganites exhibit maximum *MR* values at insulator-metal (*IM*) transition which coexist with ferromagnetic (*FM*)-paramagnetic (*PM*) transition. The observation of the *CMR* at low temperatures and under high magnetic fields (~7T) has restricted their practical usage so far. Efforts are being carried out to obtain sizeable *MR* values at low fields (*LFMR*) and near room temperature [9-10]. In polycrystalline manganites, occurrence of *LFMR* due to spin polarized tunneling [11] or spin dependent scattering [12] of the carriers at the grain boundaries has been reported which makes these ceramic materials superior candidates over the single crystalline or epitaxial manganites thin film. Several manganite materials like LCBMO, LCSMO, PBMO, and their composites with $Ag_2O$ and PdO have been investigated to explore their potential for magnetic sensing applications [9, 10, 13-15]. Besides, manganites have also shown good potential candidate as active magnetic refrigerant (*AMR*) near room temperature [13, 16-21].

In the present scenario of energy sources with environment, consuming cost, availability considerations etc., new promising sources overcoming these limitations are currently being probed into. Refrigeration by conventional gas compression technique has several limitations which can be overcome by using magnetic refrigeration [22].

Magnetic refrigeration is based on principle of magnetocaloric effect (*MCE*) involving magnetization and demagnetization of the magnetic material and its measure is given by isothermal magnetic entropy change ($\Delta S_M$) or adiabatic temperature change ($\Delta T_{ad}$) with applied magnetic field. For considerable *MCE*, $|\Delta S_M|$ should be large and constant over working temperature range. For application purposes, high *MCE* value with low applied field at near room temperature is desired. Some materials like Gd have shown potential as active magnetic refrigerants but are quite expensive [23]. Since manganites can be used as *AMR*, several attempts have been made in which Pr-based manganites can be used near room temperature *AMR* [21]. However, the value of $|\Delta S_M|$ is not much impressive for household applications and needs to be further enhanced.

There have been only a few reports about the $Pr_{2/3}Sr_{1/3}MnO_3$ material, having both the insulator-metal and Curie transitions close to room temperature [16, 24]. Moreover, there seems to be no systematic report existing to investigate or optimize its magnetic properties with respect to the sintering temperature. In particular, we consider here Pr based manganite ($Pr_{2/3}Sr_{1/3}MnO_3$) having favorable properties such as large magnetic moment, sharp *I-M* transition as well as ferro- to para-magnetic transition near room temperature, high *LFMR* and stability at ambient conditions which makes it a promising candidate for such application. An attempt is made to investigate the magneto-resistance and magnetocaloric properties of $Pr_{2/3}Sr_{1/3}MnO_3$ material vis-à-vis the sintering temperature.

## 2. Experimental

Various $Pr_{2/3}Sr_{1/3}MnO_3$ samples have been synthesized in air by the standard solid-state reaction route with starting stoichiometric mixture of $Pr_6O_{11}$, $SrCO_3$ and $MnO_2$ (all with 5N purity). These ingredients are ground thoroughly, calcined at 900ºC for 12h and pre-sintered at 1100ºC, 1200ºC for 20h with intermediate grindings. These powders are then sintered in pellet form for 24h in air at 1260ºC, 1350ºC, 1400ºC and 1450ºC respectively and finally furnace cooled to room temperature. Phase purity and the lattice parameter refining of the samples is checked through powder diffractometer using CuKα radiation (Rigaku) and the Rietveld refinement programme (Fullprof version). Scanning electron microscope (Ziess EVO MA-10) with attached energy dispersive analysis (*EDS*) was used for the surface morphology of the various samples. Resistivity and magnetization measurements of the samples are carried out in magnetic field up to 7T using a Physical Properties Measurement System (PPMS-14T, Quantum Design).

3. **Results and discussion**

Phase purity of the synthesized materials sintered at different temperatures between 1260°C-1450°C has been analysed through Rietveld refinement of their X-ray diffractograms. Fig.1(a)-(d) depict the Rietveld refined *XRD* patterns of various $Pr_{2/3}Sr_{1/3}MnO_3$ samples confirming the presence of all the identified characteristic peaks and fitted in the orthorhombic *Pbnm* space group. The fitting parameters of the Rietveld refinement and the lattice parameters are listed in the Table-I. There is no considerable change in lattice parameter with sintering temperature and the lattice volume is almost constant. Scanning electron micrographs (*SEM*) of $Pr_{2/3}Sr_{1/3}MnO_3$ samples sintered at 1260°C, 1350°C, 1400°C and 1450°C are shown in Fig. 2(a)-(d) respectively. For $T_s$ = $1260^0C$, grain size is found to be in 1-3 μm range and it is increased to 4-7 μm for $T_s$ = $1350^0C$. Sintering temperature can be noticed to have a marked enhancement of the grain growth with the samples with $T_s$ = 1400°C and 1450°C almost exhibiting single grain nature, [Fig.2 (c) and (d)]. Higher sintering temperatures have been chosen with a view to reach to nearly its melting point where the grain boundary effect would be rendered negligible. It is worthwhile mentioning here that the stoichiometry of the material is kept intact even in such a scenario as may be seen from Fig. 1.

The normalised resistance of the samples are shown in Fig. 3. The inset shows the corresponding *TCR* of the samples. Remarkable sharpness in the insulator-metal transition ($T^{IM}$) is observed with increasing $T_s$. *TCR* is found to increase from 0.52% ($T_s$ = $1260^0$C) to 8.4% ($T_s$ = $1450^0$C). The insulator-metal transition ($T^{IM}$) is observed close to room temperature. Temperature dependence of the magneto-resistance *MR(T)* of the samples at magnetic field values of 1T and 5T is depicted in Fig. 4. Interestingly, *MR* data clearly shows an increasing trend in the Peak *MR* value. Material sintered at 1400°C exhibits *MR* as high as 32% at 1T and 58% at 5T. Even though the 1450°C material shows higher peak *MR* value at 5T field, slightly lower value of 58% obtained in the 1400C material at higher temperature of ~284K is clearly advantageous. This feature is better depicted in *MR* data at 1T field value.

The magnetization vs temperature [*M(T)*] graph for 500 Oe of applied magnetic field shows increased sharpness of transition with $T_s$ [Fig. 5]. The sharpness can also be seen by *dM/dT* vs temperature graph in inset of *M(T)* graph. The Curie temperature of the samples is observed around 284K which is calculated by taking the derivative of magnetization with temperature. Slight shift in $T_c$ towards lower temperature is also observed with increasing $T_s$.

To measure the magnetocaloric effect one can employ the route of measuring the adiabatic change of temperature with magnetic field or through isothermal magnetization versus field at different temperatures [13, 25]. In the present study the isothermal magnetization measurement route with the applied field up to 5 T has been adopted. The change in magneto-entropy is related to the magnetization by the Maxwell relation and details can be found in refs [13, 25] wherein,

$$\Delta S_M(T,H) = \int_0^{H_m} \left(\frac{\partial M}{\partial T}\right)_H dH$$

The magneto-entropy change can be obtained through isothermal magnetization data measured at small temperature intervals and discrete fields. Isothermal magnetization curves for all the sintering temperature are shown in Fig. 6. (a). In this study isothermal curves have been obtained around $T_c$ from 271K to 300K at close intervals of 1K near $T_C$. For better understanding of $M(T)$ and isothermal magnetization data we have drawn Arrott plots. These plots help to understand the order of transition, depicted in Fig. 6 (b). Arrott plots suggest that the transition is nearly first order (discontinuous), which is tending more towards it with increasing $T_s$. As sharpness of the magnetic transition increases, the value of $|\Delta S_M|$ increases. That is why increased $|\Delta S_M|$ values are observed with increasing $T_s$.

Fig. 7 shows the temperature dependence of the magnetic entropy change of $Pr_{2/3}Sr_{1/3}MnO_3$ samples with $T_s$ 1260 and 1400$^0$C at field changes of 1 to 5 T. In the same magnetic field change, $|\Delta S_M|$ variation with temperature shows a peak near $T_c$ (284K). Upon the change of 5 Tesla applied field, the highest value of $|\Delta S_M|$ is 7.8 J.Kg$^{-1}$.K$^{-1}$. This compares excellently with the value of $|\Delta S_M|$ is 7.45 J.Kg$^{-1}$.K$^{-1}$ with the same change of 5T field reported in single crystal of $La_{0.7}Ca_{0.2}Sr_{0.10}MnO_3$ [18]. The is higher than −1.75 J.Kg$^{-1}$.K$^{-1}$ obtained at $T_c$ = 260 K in $Pr_{0.7}Sr_{0.3}MnO_3$ [26](EPL). The $|\Delta S_M|$ value obtained at 4T is 6.5 J.Kg$^{-1}$.K$^{-1}$ is higher than the obtained value 3.88 J.Kg$^{-1}$.K$^{-1}$ at 4.5T at 334 K in La$_{0.6}$Sr$_{0.2}$Ba$_{0.2-x}$□$_x$MnO$_3$ [27](JAP). However, the same is lower than the value of 8.52 J.Kg$^{-1}$.K$^{-1}$ in single crystal of Pr$_{0.63}$Sr$_{0.37}$MnO$_3$ [16]. Further PSMO with higher sintering temperature also seems to have an edge over composite LCSMO:Ag manganite as well [13]. It is also evident that the synthesized material not only consist remarkable magnetic entropy value but also high TCR close to room temperature. Double peaks in $|\Delta S_M|$-$T$ graph have been observed with all samples invariant of $T_s$. Reports about similar nature also been observed in magneto-caloric compounds such as Tb$_5$Gb$_4$ [28], La$_{0.67}$Sr$_{0.33}$Mn$_{0.9}$Cr$_{0.1}$O$_3$ [20]. The double peak behavior is explained by considering presence of two types of ferromagnetic structures/clusters, inhomogeneity in stoichiometry, complex magnetic structure originated from dopant [20, 28] and grain boundary effects [19]. In the studied samples the grain boundary effect can be discarded as both the peaks are present in all samples. It is possibly occurring due to complex magnetic structure caused by variable valence state of Pr ions. It is expected that the *MCE* range should be wide so that the working temperature range, which is required in Ericson cycle. The occurrence of double peaks can be utilized to obtain wide range of working temperature $\Delta T$. The underline philosophy in going through higher sintering temperatures has been to reduce or remove the effect of grain boundary density which seems to have resulted in the improved MCE values. This clearly has led to the reduction of the resistive intergrain regions. Significantly, the material also exhibits a high TCR value of 8.4% around its $T_c$. Therefore, such Pr$_{2/3}$Sr$_{1/3}$MnO$_3$ sample having good MR (32% 1T, 58% 5T) and reasonable change in magnetic entropy (7.8Jkg$^{-1}$K$^{-1}$, 5T) generates its possibility of a useful magnetic sensing and refrigerant material close to room temperature.

**Conclusions**

Improvement in both the magnetic (*MR*) and magnetocaloric (*MCE*) properties of $Pr_{2/3}Sr_{1/3}MnO_3$ manganite material has been observed as a function of the final sintering temperature $T_s$. Both the insulator-metal transition ($T^{IM}$) and Curie temperatures ($T_c$) are observed close to room temperature. *MR* and TCR are found to increase with $T_s$. Material sintered at 1400°C exhibits *MR* as high as 32% at 1T and 58% at 5T. The maximum entropy change ($\Delta S_M max$) near its $T_C$ is 2.3 Jkg$^{-1}$K$^{-1}$ and 7.8 Jkg$^{-1}$K$^{-1}$ at 1T and 5T fields respectively. Such a $Pr_{2/3}Sr_{1/3}MnO_3$ material having a sizeable *MR* and reasonable change in magnetic entropy shows promise of an active magnetic sensor and magnetic refrigerant close to room temperature.

**Acknowledgements**

Authors thank the Director, National Physical Laboratory, New Delhi, for his support and encouragement. The authors SKS and RCB would like to thank the Council of Scientific and Industrial Research (CSIR), New Delhi for the Senior Research Fellowship and the financial assistance under the Emeritus Scientist Scheme (Dr. S.K. Agarwal) respectively.

**Table and Figures**

**Table 1** Rietveld refined lattice parameters and unit cell volume of $Pr_{2/3}Sr_{1/3}MnO_3$ samples sintered at different temperatures.

**Table 2** Insulator-Metal transition temperature $T_{IM}$, Maximum TCR values and Ferromagnetic to Paramagnetic transition temperature $T_c$ of $Pr_{2/3}Sr_{1/3}MnO_3$ samples sintered at different temperatures.

**Fig. 1.** Reitveld refined *XRD* patterns of various $Pr_{2/3}Sr_{1/3}MnO_3$ samples.

**Fig. 2.** Scanning electron micrographs of $Pr_{2/3}Sr_{1/3}MnO_3$ samples sintered at

a) 1260C, b) 1350C, c) 1400C, d) 1450C.

**Fig. 3.** Normalized Resistance versus temperature for various $Pr_{2/3}Sr_{1/3}MnO_3$. Inset shows the correponding *TCR* of the samples.

**Fig. 4.** MR versus temperature for various $Pr_{2/3}Sr_{1/3}MnO_3$ samples at $H = 1$ T and 5 T

**Fig. 5.** DC magnetization versus temperature for various $Pr_{2/3}Sr_{1/3}MnO_3$. Inset shows the correponding *dM/dT* of the samples.

**Fig. 6. (a)** Isothermal magnetization *M(H)* at close *T* intervals and **(b)** Arrott plots for various $Pr_{2/3}Sr_{1/3}MnO_3$.

**Fig. 7.** Temperature dependence of entropy change of $Pr_{2/3}Sr_{1/3}MnO_3$ ($T_s$ = 1260$^0$C and 1400$^0$C) samples at $H$ = 1 T, 2T, 3T, 4T and 5 T fields.

**Table 1**

| $Pr_{2/3}Sr_{1/3}MnO_3$ | a (Å) | b (Å) | c (Å) | Vol. (Å$^3$) |
|---|---|---|---|---|
| $T_S$ = 1260 $^0$C | 5.44 (3) | 5.46 (1) | 7.72 (5) | 229.65 (5) |
| $T_S$ = 1350 $^0$C | 5.45 (5) | 5.45 (2) | 7.72 (3) | 229.72 (3) |
| $T_S$ = 1400 $^0$C | 5.44 (7) | 5.45 (3) | 7.72 (1) | 229.41 (9) |
| $T_S$ = 1450 $^0$C | 5.44 (8) | 5.44 (5) | 7.72 (1) | 229.07 (1) |

**Table 2**

| $Pr_{2/3}Sr_{1/3}MnO_3$ | Insulator-Metal Transition Temperature ($T_{IM}$) (K) | Maximum Temperature Coefficient of Resistance ($TCR_{MAX}$) (%) | FM-PM Transition Temperature (K) |
|---|---|---|---|
| $T_S$ = 1260 $^0$C | 286.0 | 0.5 | 279.5 |
| $T_S$ = 1350 $^0$C | 287.5 | 2.3 | 283.5 |
| $T_S$ = 1400 $^0$C | 295.5 | 7.1 | 280.5 |
| $T_S$ = 1450 $^0$C | 297.0 | 8.4 | 280.0 |

**Fig. 1**

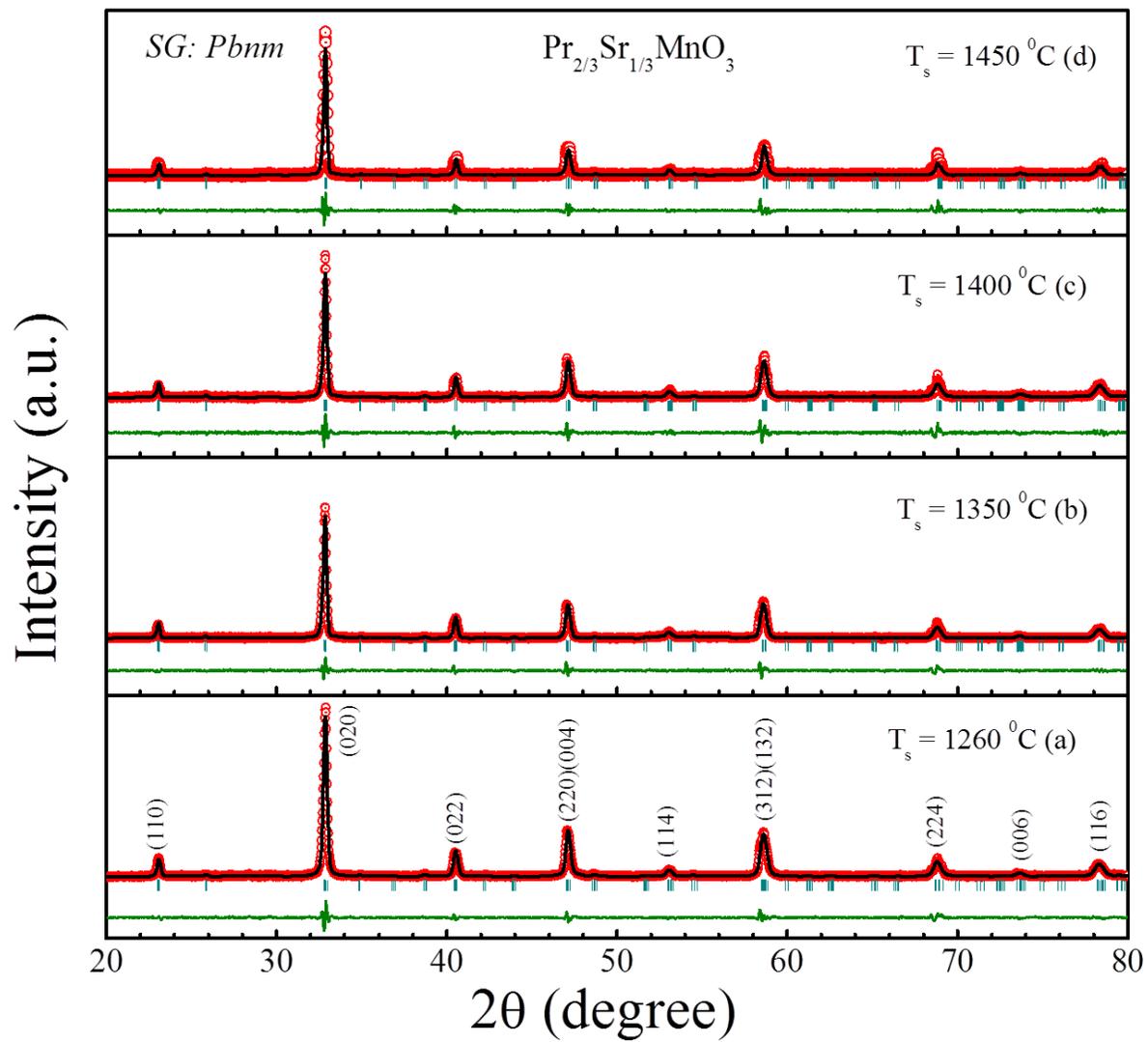

**Fig. 2**

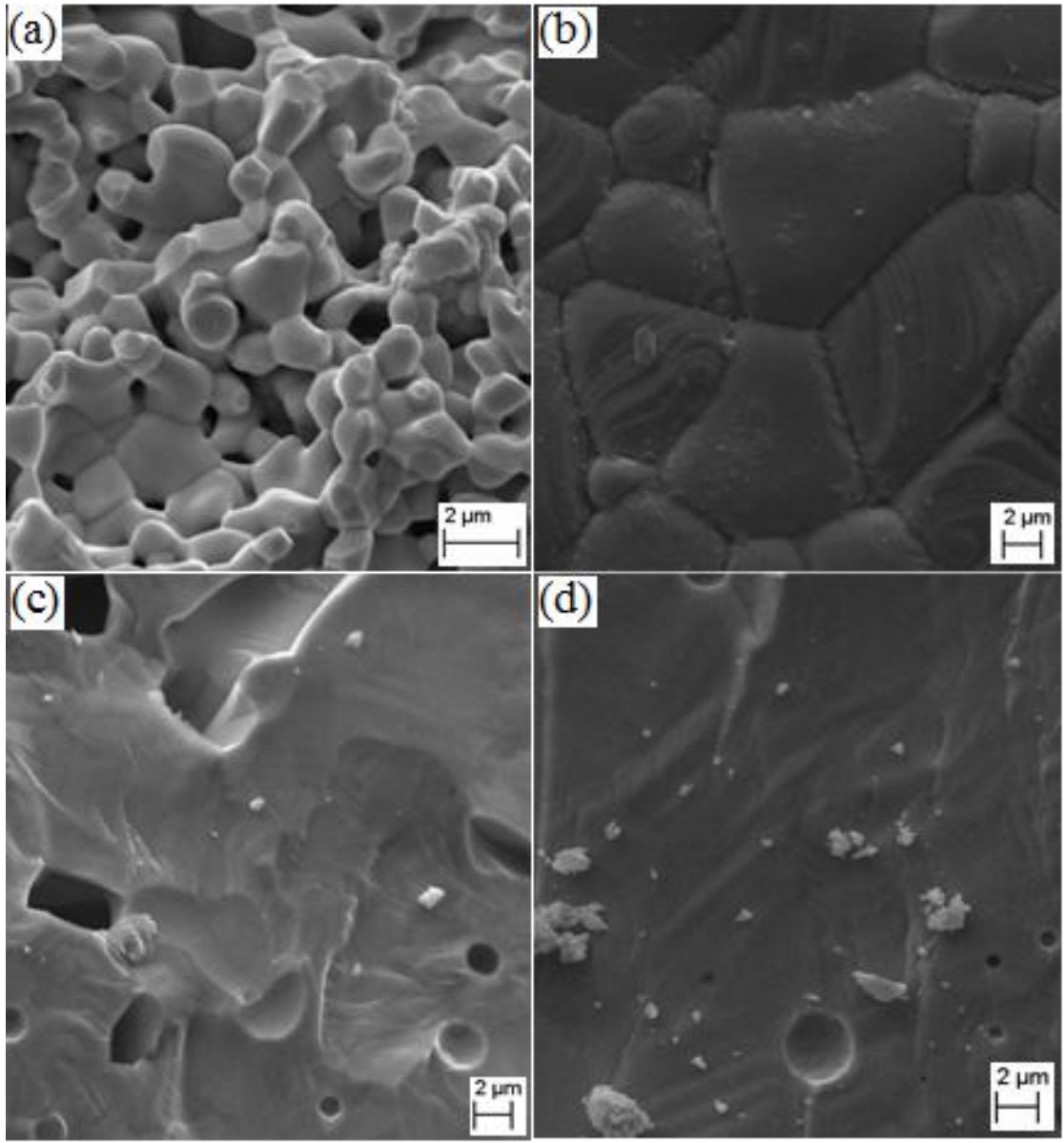

**Fig. 3**

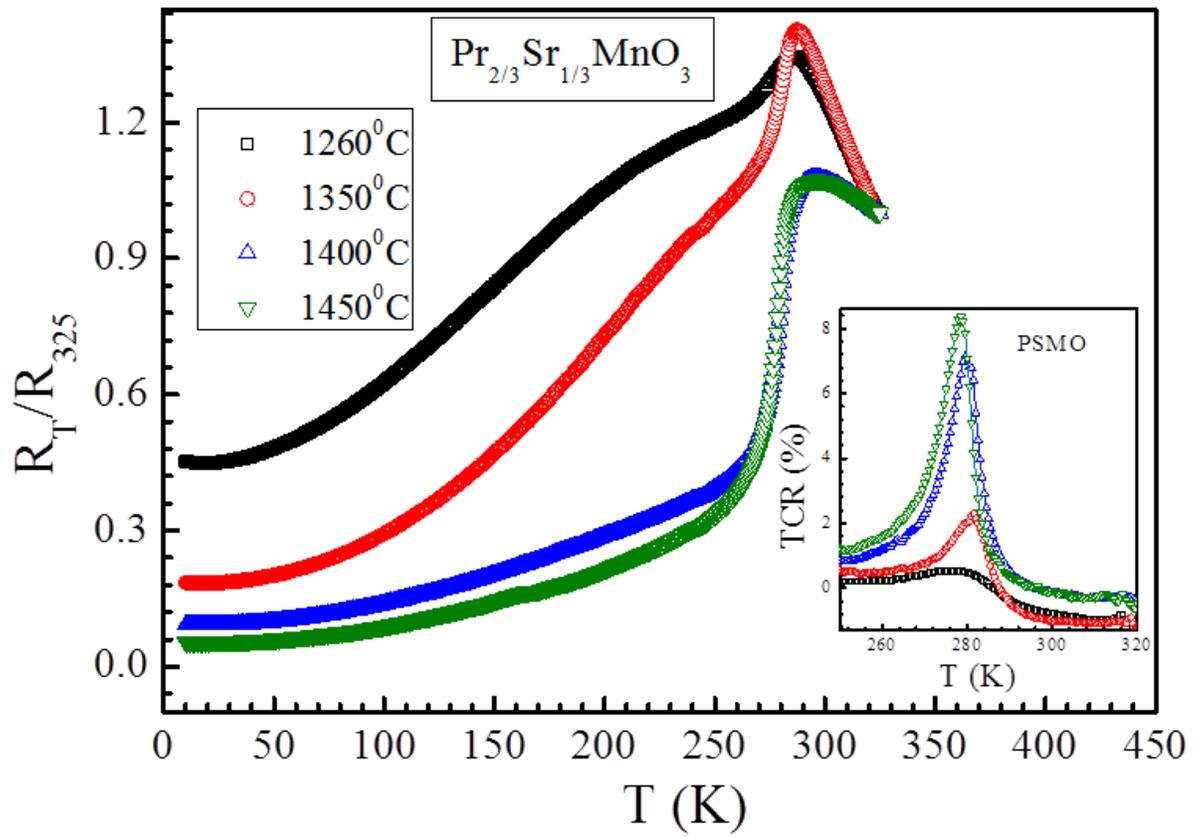

**Fig. 4**

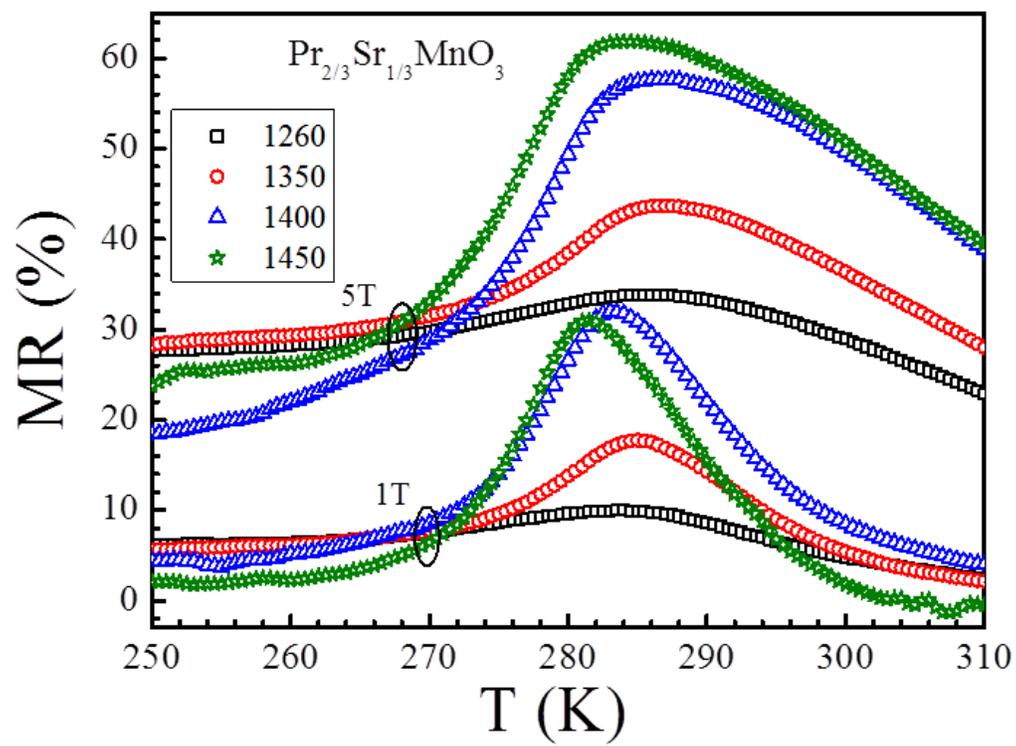

**Fig. 5**

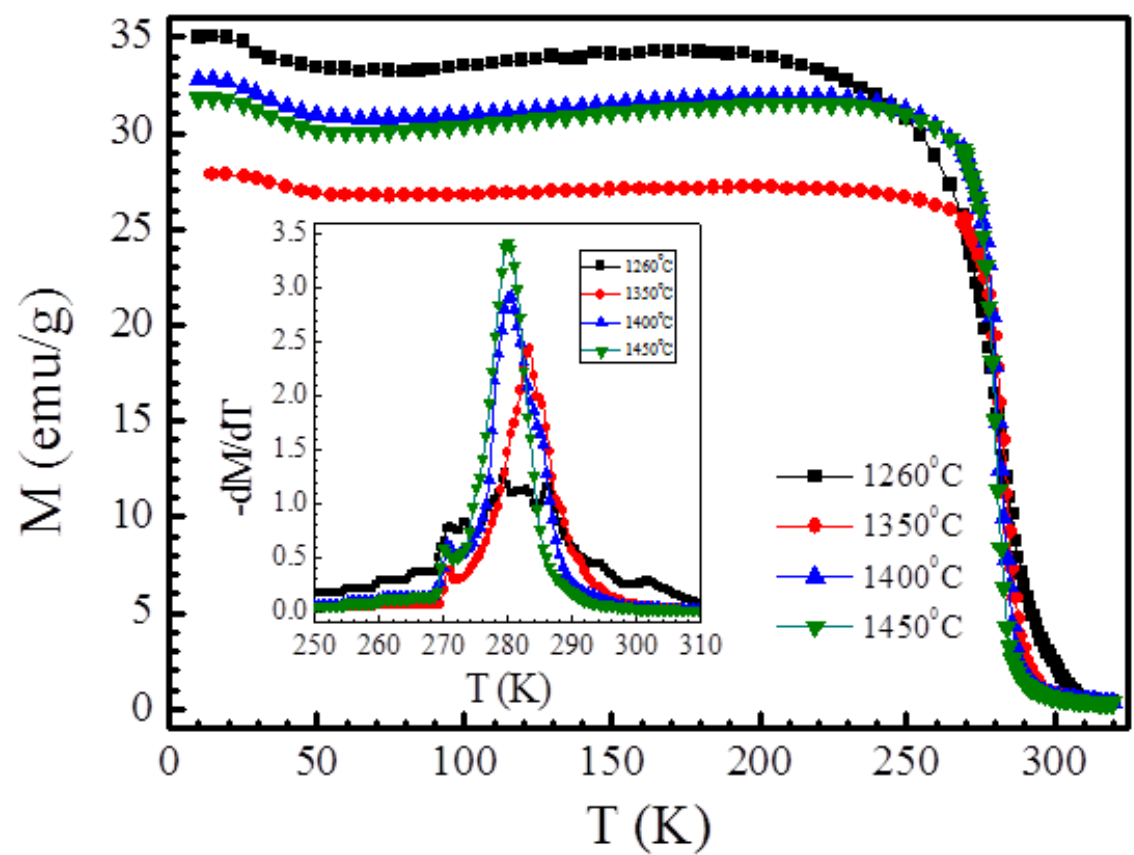

**Fig. 6(a)**                    **Fig. 6(b)**

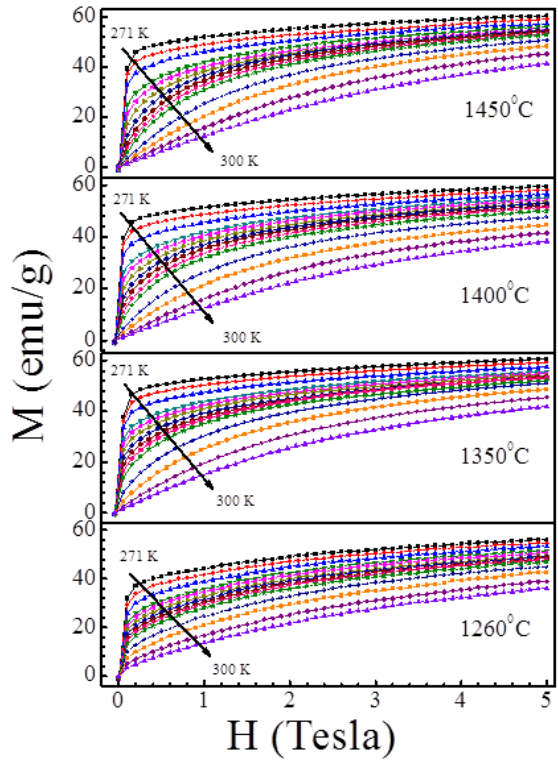
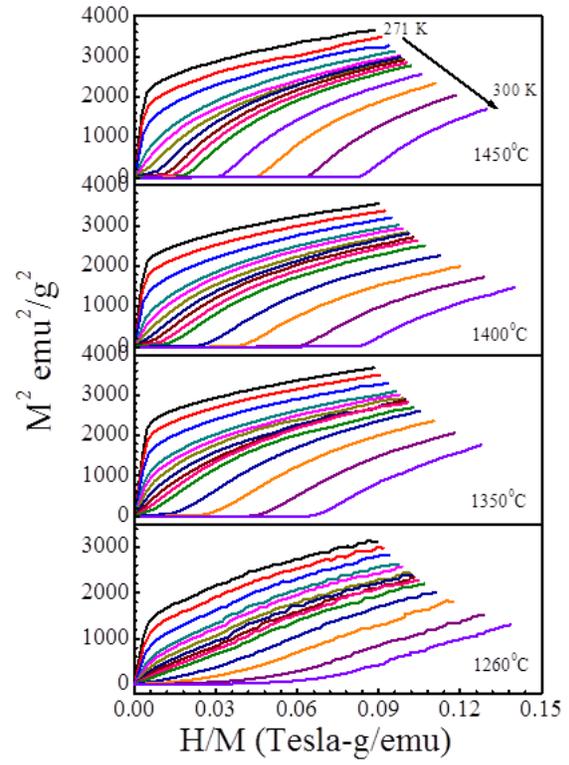

**Fig. 7**

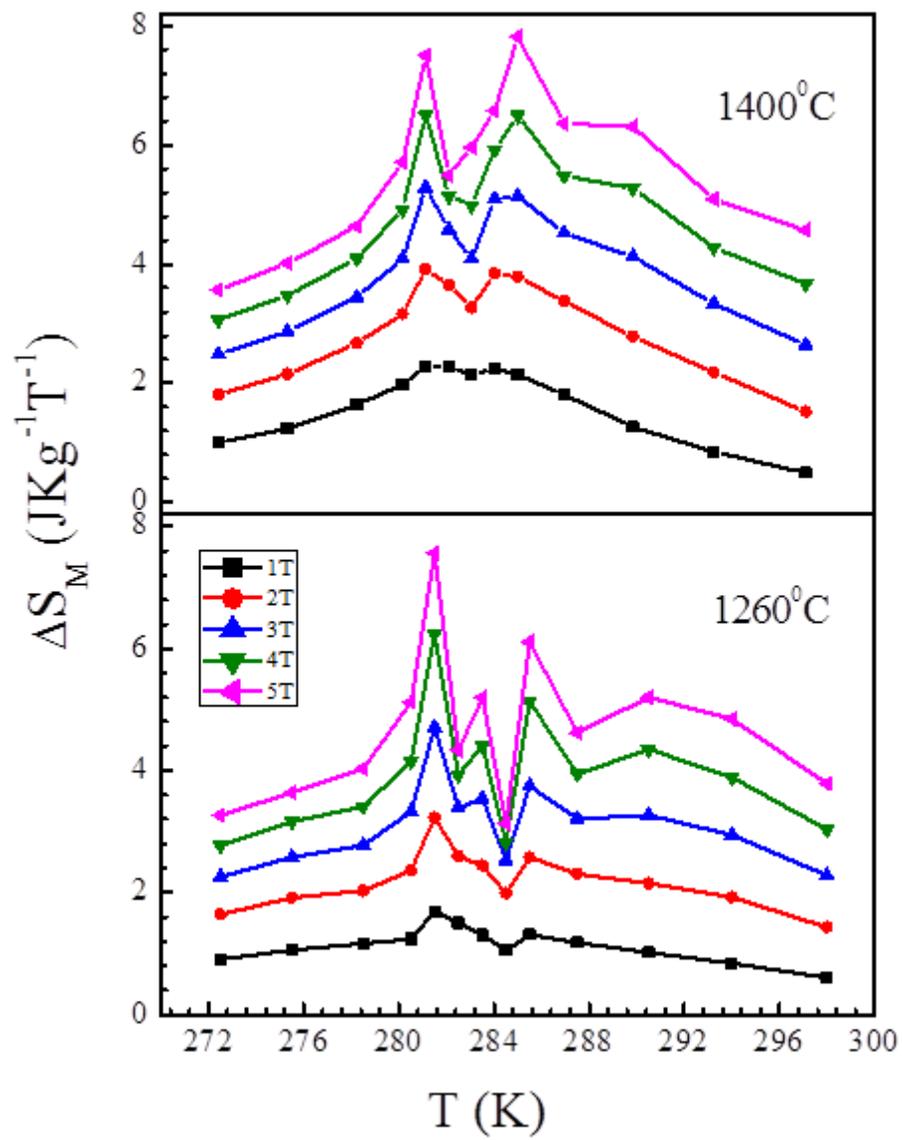